
\def\chaphead{}
\def\ni{\noindent}

\font\tfont=cmbxti10
\font\eightrm=cmr8
\font\eightit=cmti8
\font\sixrm=cmr6
\font\eightmit=cmmi8
\font\sixmit=cmmi6
\def\absmath{\textfont0=\eightrm \scriptfont0=\sixrm
	      \textfont1=\eightmit \scriptfont1=\sixmit}
\def\absfont{\let\rm=\eightrm \let\it=\eightit \rm\absmath}
\font\twelverm=cmr12
\font\twelveit=cmti12
\font\tenrm=cmr10
\font\twelvemit=cmmi12
\font\tenmit=cmmi10
\def\regmath{\textfont0=\twelverm \scriptfont0=\tenrm
	      \textfont1=\twelvemit \scriptfont1=\tenmit}
\def\peterfont{\let\rm=\twelverm \let\it=\twelveit \rm\regmath}
%
%

\newfam\vecfam

\textfont\vecfam=\tfont \scriptfont\vecfam=\seveni
\scriptscriptfont\vecfam=\fivei


\def\spose#1{\hbox to 0pt{#1\hss}}

\font\eightrm=cmr8

\def\s{\ifmmode \widetilde \else \~\fi} 
     
\def\section{\S}
\newcount\notenumber
\notenumber=1
\newcount\eqnumber
\eqnumber=1
\newcount\fignumber
\fignumber=1
\newbox\abstr


\def\s{{\rm\,s}}

\def\note#1{\footnote{$^{\the\notenumber}$}{#1}\global\advance\notenumber by 1}
\def\foot#1{\raise3pt\hbox{\eightrm \the\notenumber}
     \hfil\par\vskip3pt\hrule\vskip6pt
     \noindent\raise3pt\hbox{\eightrm \the\notenumber}
     #1\par\vskip6pt\hrule\vskip3pt\noindent\global\advance\notenumber by 1}

\def\abstract#1{\setbox\abstr=\vbox{\hsize 5.0truein{\par\noindent#1}}
    \centerline{ABSTRACT} \vskip12pt \hbox to \hsize{\hfill\box\abstr\hfill}}
     
\def\Dt{\spose{\raise 1.5ex\hbox{\hskip3pt$\mathchar"201$}}}    
\def\dt{\spose{\raise 1.0ex\hbox{\hskip2pt$\mathchar"201$}}}    

\def\new{{\rm\chaphead\the\eqnumber}\global\advance\eqnumber by 1}
\def\ref#1{\advance\eqnumber by -#1 \chaphead\the\eqnumber
     \advance\eqnumber by #1 }
\def\last{\advance\eqnumber by -1 {\rm\chaphead\the\eqnumber}\advance
     \eqnumber by 1}
\def\eqnam#1{\xdef#1{\chaphead\the\eqnumber}}
     
\def\nfig{\chaphead\the\fignumber\global\advance\fignumber by 1}
\def\nfiga#1{\chaphead\the\fignumber{#1}\global\advance\fignumber by 1}
\def\rfig#1{\advance\fignumber by -#1 \chaphead\the\fignumber
     \advance\fignumber by #1}
\def\fignam#1{\xdef#1{\chaphead\the\fignumber}}

\def\lta{\mathrel{\spose{\lower 3pt\hbox{$\mathchar"218$}}
     \raise 2.0pt\hbox{$\mathchar"13C$}}}
\def\gta{\mathrel{\spose{\lower 3pt\hbox{$\mathchar"218$}}
     \raise 2.0pt\hbox{$\mathchar"13E$}}}
     

\magnification=\magstep1
\parskip=3pt

\magnification=\magstep1
\def\ni{\noindent}
\def\spose#1{\hbox to 0pt{#1\hss}}
\def\lta{\mathrel{\spose{\lower 3pt\hbox{$\mathchar"218$}}
     \raise 2.0pt\hbox{$\mathchar"13C$}}}
\def\gta{\mathrel{\spose{\lower 3pt\hbox{$\mathchar"218$}}
     \raise 2.0pt\hbox{$\mathchar"13E$}}}

\def\frac#1#2{{#1\over #2}}
\baselineskip=14pt

\centerline{\bf Angular momentum transport by gravity waves}
\centerline{\bf and its effect on the rotation of the solar interior}

\vskip 0.7truecm
\centerline{Pawan Kumar$^\dagger$}
\centerline{Institute for Advanced Study, Princeton, NJ 08540}
\medskip
\centerline{and}
\medskip
\centerline{Eliot J. Quataert}
\centerline{Harvard-Smithsonian Center for Astrophysics, 60 Garden St., 
Cambridge, MA 02138}

\bigskip\bigskip
\centerline{\bf Abstract}

We calculate the excitation of low frequency gravity waves by
turbulent convection in the sun and the effect of the angular momentum
carried by these waves on the rotation profile of the sun's radiative
interior. We find that the gravity waves generated by convection in
the sun provide a very efficient means of coupling the rotation in the
radiative interior to that of the convection zone.  In a
differentially rotating star, waves of different azimuthal number have
their frequencies in the local rest frame of the star Doppler shifted
by different amounts. This leads to a difference in their local
dissipation rate and hence a redistribution of angular momentum in the
star. We find that the time scale for establishing uniform rotation
throughout much of the radiative interior of the sun is $\sim 10^7$
years, which provides a possible explanation for the helioseismic
observations that the solar interior is rotating as a solid body.

\ni{\it Subject headings:} waves --- angular momentum --- stars: rotation

\vskip 1.5truecm
\noindent $^\dagger$Alfred P. Sloan Fellow \& NSF Young Investigator

\noindent email addresses: pk@sns.ias.edu (P. Kumar)

\hskip 55pt equataert@cfa.harvard.edu (E.J. Quataert)

\vfill\eject
\centerline{\bf 1. Introduction}
\bigskip

Our understanding of the internal structure of the sun has improved
significantly since the advent of helioseismic mapping.  The new
dedicated observing programs such as GONG and SOHO are providing a map
of the sun's internal structure with increasing resolution and
accuracy (Christensen-Dalsgaard et al. 1996).  We already have an
excellent measurement of the thickness of the convection zone
(Christensen-Dalsgaard et al. 1996), and the rotation profile in the
outer half of the sun (Thompson et al. 1996).  These results suggest
that the radiative interior of the sun is rotating as a solid body at
a rate equal to the rotation rate of the base of the convection zone.
It has been suggested that magnetic torques in the radiative interior
of the sun might be responsible for its solid body rotation. In this
paper, we show that gravity waves generated by turbulent stresses in
the convection zone carry significant angular momentum and are very
efficient in establishing solid body rotation throughout much of the
radiative interior.

The calculation of the mechanical energy luminosity in gravity waves
due to turbulent stresses in the convection zone is discussed in the
next section. The angular momentum luminosity in gravity waves and its
effect on the rotation profile of the solar interior is discussed in
section 3.

\bigskip
\centerline{\bf 2. Energy luminosity in low frequency gravity waves due to turbulent 
excitation}
\bigskip

We follow the work of Goldreich, Murray, and Kumar (1994) in order to
calculate the energy luminosity in gravity waves due to excitation by
the fluctuating Reynold's stress. For simplicity, we assume that the
wave excitation is independent of the azimuthal wavenumber. The
mechanical energy luminosity in gravity waves per unit frequency,
i.e., the energy flowing through a sphere per unit time and frequency,
just below the convection zone is given by

$$L_E^{(c)}(\omega, \ell) = \omega^2\int dr\, r^2 \rho^2
\left[\left(\frac{\partial
\xi_r}{\partial r}\right)^2 + \ell(\ell+1)\left(\frac{\partial 
\xi_h}{\partial r}\right)^2 \right]
\frac{v^3 L^4 \zeta^2}{1 + (\omega T)^{15/2}}, \eqno(\new)$$  
where $\xi_r$ and $\xi_h$ are the radial and horizontal displacement
wavefunctions which are normalized to unit energy flux just outside
the convection zone, $\ell$ is the spherical harmonic degree of the
wave, $v$ is the convective velocity, $L$ is the radial
size of an energy bearing turbulent eddy, $T \approx L/v$ is the
characteristic convective time, and $\zeta$ is the ratio of the horizontal
and vertical length scales of the turbulent eddies.

For turbulence in the solar convection zone, $L \sim \alpha H$ and
$\zeta \sim 1$, where $\alpha \sim 1.65$ is the mixing length
parameter and $H$ is the pressure scale height. Gravity waves of
long horizontal wavelength, i.e. low $\ell$, and 
frequency $\omega$ are most efficiently excited by energy bearing
turbulent eddies which have turnover times comparable to
$\omega^{-1}$.  The energy luminosity per unit frequency, as seen by an
observer corotating with the base of the convection zone, at
frequencies greater than the characteristic convective frequency at
the base of the convection zone ($N_c \approx 0.15 \mu$Hz), is readily
estimated from equation (1): $$L_E^{(c)}(\omega,
\ell)
\approx \frac{[\ell(\ell+1)]^{3/2} L_\odot}{\omega} \left(\frac{H_w}{r_w}
\right)^5 \left(\frac{r_\omega}{r_c}\right)^{-2\sqrt{\ell(\ell+1)}} 
\approx \frac{10^{29}}{N_c}
\left(\frac{N_c}{\omega}\right)^{13/3} \ {\rm erg}, 
\eqno(\new)$$ where
$L_\odot \approx 4 \times 10^{33}$ erg s$^{-1}$ is the solar
luminosity and the subscript $\omega$ indicates a quantity evaluated
at $r$ such that $\omega T(r) \sim \pi/2$.  The latter equality in
equation (2) is only applicable to dipole waves and follows from using
$r_\omega \approx r_c \approx 0.72 R_\odot$ and $H_\omega \approx H_c
(\omega/N_c)^{-3/(n+3)}$, where $H_c \approx 0.1r_c$ is the pressure
scale height at the base of the convection zone and $n \approx 1.5$ is
the effective polytropic index of the solar convection zone.  This is
in good agreement with the numerical evaluation of equation (1) which
yields a power law of index $-4.5$.  For $\omega \approx N_c$, our
estimate of $L^{(c)}_E$ is, to within a factor of a
few, equivalent to that of Press (1981), who considered the effect of
low frequency gravity waves on elemental mixing in the radiative
interior of the sun.  For $\omega \gta N_c$, however, Press
underestimated the energy input into gravity waves by only considering
the excitation due to inertial range eddies at $r \approx r_c$; the
Kolmogorov scaling gives $L_E^{(c)} \propto \omega^{-8.5}$ for
inertial range eddies.\footnote{$^1$}{Waves with $\omega \gta N_c$ also
get excited by turbulence in the convective overshoot layer.  If the
thickness of the overshoot layer is $\sim \eta H_c$ then, from
equation (1), we estimate that the luminosity generated at $\omega \sim
N_c \eta^{-1}$ by turbulence in the convective overshoot layer is
$\sim \eta^{-4/3}$ times larger than that generated at the same
frequency in the convection zone. We note that the singularity
as $\eta \rightarrow 0$ does not apply to stars since 
$\omega$ for gravity waves has an upper bound of order the maximum
Brunt-V\"ais\"al\"a~frequency in the star.}

\bigskip 
\centerline{\bf 3. Angular momentum transport by gravity waves}
\medskip

The angular momentum luminosity per unit frequency (angular momentum
flowing across a sphere of radius $r$ per unit time and frequency),
$L'_{a}(\omega, \ell, m,r)$, associated with a gravity wave of
frequency $\omega$ and azimuthal order $m$ can be shown to be equal to
(e.g., Goldreich \& Nicholson 1989) $$ L'_{a}(\omega,\ell, m,r) = {-m
L_E(\omega,\ell,m,r)\over
\omega},\eqno(\new)$$ where $L_E$ is the mechanical energy luminosity
per unit frequency associated with the wave.  It should be pointed out
that $L_E$ and $\omega$ are
frame dependent quantities, but the ratio $L_E/\omega$, the
luminosity of wave-action, is frame independent. Thus the value of
$L'_a$ does not vary as the wave propagates through a non-dissipative
shearing medium.  We find it convenient to
evaluate the frame dependent quantities, $L_E$ \& $\omega$, as seen by
an observer corotating with the base of the convection zone (where the
waves are generated). The net angular momentum luminosity in dipole
gravity waves of a non-zero $m$ at the base of the convection zone is
obtained by combining equations (2) and (\ref1) and is $\sim 10^{35}$
dyne cm.  This luminosity, deposited in the sun over $\approx$ 10$^7$
years, is equal to the total angular momentum in the solar interior
(provided that waves of opposite $m$ values do not deposit their
angular momentum at the same location, which we show below is the
case). Thus gravity waves should have a significant effect on the
interior rotation profile of the sun.

Differential rotation along latitudes in radiative stars is subject to 
instabilities which restore uniform rotation (e.g.  Balbus \& Hawley 1996)
and so we consider the angular rotation frequency in the radiative
interior of the sun, $\Omega$, 
to be a function of $r$ alone. The rotation speed at the base of the
convection zone is denoted by $\Omega_c$. The wave frequency in the
rest frame of the fluid at radius $r$, $\omega^{(m)}_*(r)$, is given
below in the terms of its frequency $\omega$ at the base of the
convection zone 
$$ \omega^{(m)}_*(r) = \omega + m[\Omega_c - \Omega(r)]. \eqno(\new)$$ 
If we ignore wave dissipation and assume that there are
no corotation resonances in the system, i.e., $\omega^{(m)}_*(r)$ is
non-zero everywhere, then the net angular momentum luminosity carried
by gravity waves is zero everywhere in the radiative interior and
therefore waves do not modify the rotation profile of the sun.

However, the radiative dissipation rate of gravity waves
depends on their frequency in the local rest frame of the fluid, which
is different for waves of $\pm m$ in a differentially rotating star.
This leads to a non-zero net angular momentum
luminosity whenever the rotation profile of the star deviates from
solid body rotation.

The local radiative dissipation rate for gravity waves can be
expressed as $$\gamma (\omega, \ell, r) \approx \frac{F_r k^2_r}{\rho
c_s^2 d lnT/dr} \left(\frac{\partial lnT}{\partial ln \rho} \right)_p
\left(\frac{\partial lnT}{\partial ln \rho} \right)_S
\approx {2 F_r k_r^2 H_T\over 5 p} ,\eqno(\new)$$ 
where $k_r\approx N\sqrt{\ell(\ell+1)}/[r \omega^{(m)}_*(r)]$ is the
wave's radial wave number in the local rest frame of the fluid, $F_r$
is the radiative flux a distance $r$ from the center, $c_s$ is the
sound speed, $p$ is the pressure and $H_T$ is the temperature scale
height.

~From equation (\last) we see that low frequency gravity waves with
$\omega \approx N_c$ are strongly damped near the base of the solar
convection zone.  For the solar model we are using (due to
Christensen-Dalsgaard) the damping length of dipole waves with $\omega
\approx N_c$ is only $\sim 10$ wavelengths.  Waves with $\omega \gta N_c$ 
carry less energy, but their damping length is longer and so they
propagate deeper into the radiative interior.  The energy luminosity
per unit frequency in gravity waves of frequency $\omega$, degree
$\ell$, and azimuthal order $m$, a distance $r$ from the center, is given by:
$$ L_E(\omega,\ell,m,r) = L_E^{(c)}(\omega,\ell)
\exp[-\tau(\omega,\ell,m,r)], \eqno( \new)$$ where 
$$ \tau(\omega,\ell,m,r) \equiv \int_{r}^{r_c} dr'\,
{\gamma(\omega_*^{(m)}(r'),
\ell,r')   \over v_g(\omega_*^{(m)}(r'),\ell,r')} \eqno(\new )$$ is
the attenuation depth of the wave and $v_g \approx
\omega^{(m)}_*(r)/k_r$ is the group speed of gravity waves in the radial
direction.  We note that $\gamma/v_g \propto \omega^{-4}
[\ell(\ell+1)]^{3/2}$.  The net energy luminosity is obtained by
integrating $L_E(\omega, \ell,m,r)$ over all frequencies and adding
the contribution from waves of different $\ell$ and $m$.  For the case
of solid body rotation, the net energy luminosity in the sun is shown
in Figure 1, which also shows the frequency for which $\tau \approx 1$
for dipole waves. Since wave excitation falls off with increasing
frequency and $\ell$, and the attenuation depth decreases as $\sim
\ell^3\omega^{-4}$, most of the contribution to the luminosity comes
from dipole gravity waves with frequencies such that $\tau\approx 1$.

The net angular momentum luminosity at $r$, $L_a$, is obtained by adding
waves of different $m$ \& $\ell$, and integrating over wave frequency 

$$ L_{a}(r) = \sum_{\ell, |m|} \int_0^\infty d\omega\, 
  {|m| L_E^{(c)}(\omega,\ell) \over\omega}\Biggl( \exp
  \bigl[-\tau(\omega,\ell,-|m|,r)\bigr] - 
\exp\bigl[-\tau(\omega, \ell,|m|,r)\bigr]\Biggr).   \eqno(\new)$$
In the limit of small differential rotation, $\bigl|\Delta\Omega(r)\bigr|
\equiv \bigl|\Omega(r)-\Omega_c\bigr|\ll\omega$, this equation simplifies to

$$ L_{a}(r) \approx \sum_{\ell, |m|} \int_0^\infty d\omega\, {2 |m|
L_E(\omega,\ell,0,r)\over\omega} \sinh(|m|\delta\tau), \eqno(\new)$$
where $$ \delta\tau = 4 \int_r^{r_c} dr'\;
{\gamma(\omega,\ell,r')\over v_g(\omega,\ell,r')}
{\Delta\Omega(r')\over\omega}. \eqno(\new)$$ Thus the rate at which
angular momentum is deposited in the fluid is

$$ \eqalign{-{d L_{a}(r)\over dr} \approx & -2\sum_{\ell, |m|}
\int_0^\infty d\omega\, {|m|\gamma(\omega,\ell,r)
L_E(\omega,\ell,0,r)\over \omega v_g(\omega, \ell, r)} \times \cr & \left[
4|m|\left( {\Delta\Omega(r)\over\omega}\right)\cosh(|m|\delta\tau) -
\sinh(|m| \delta\tau)\right]}. \eqno(\new)$$ Finally, the rate of
change of the rotation frequency is given by the following equation

$$ {d\over dt}\left[\Delta\Omega(r)\right] = -{3\over 8\pi\rho r^4} {d
L_a(r)\over dr}. \eqno(\new)$$ Equations (\ref2) and (\ref1) together
determine the evolution of the rotation profile of the sun. The
resulting equation is a nonlinear differentio-integral equation that
is not easy to solve.  We have, however, numerically solved the
linearized version of this equation, which is valid in the limit that
$\delta \tau \ll 1$; the result for the evolution of the angular
rotation rate of the sun is shown in Figure (2). Since $\tau\sim 1$
for the waves with most energy and since $\delta\tau<\tau
(\Delta\Omega/\omega)$, $\delta\tau< 1$ is in fact a good
approximation in the solar interior.

We now focus on providing simple estimates of the timescale over which
the rotation rate changes and a physical discussion of the angular
momentum redistribution in the sun. For small differential rotation 
($|\Delta\Omega|/\omega\ll1$) $L_a(r)$ increases linearly with $\Delta
\Omega$ and so there
is a characteristic time, $t(r)$, for the change of the angular
momentum at radius $r$ that is independent of the rotation rate $$
\left[t(r)\right]^{-1} = {3\over \pi\rho r^4} \int_0^\infty d\omega\,
 {\gamma(\omega,1,r) L_E(\omega,1,0,r)\over \omega^2 v_g(\omega,1,r)}.
\eqno(\new)$$ Figure (3) shows $t(r)$ for the sun.  Note that the
characteristic time in the radiative interior is less than $\approx$
10$^7$ years and so the rotation rate in this region is expected to be
strongly influenced by the angular momentum deposited by gravity
waves.  To gain some physical understanding of how the rotation
profile evolves, we go back to equation (9).

Note that $L_E$ is a decreasing function of $(r_c-r)$ (waves are
rapidly attenuated with distance; see Fig. 1) while
$\sinh(|m|\delta\tau$) is an increasing function.  Thus the angular
momentum luminosity, which is proportional to the product of these
functions (see eq. [9]), peaks at some depth that depends on the
rotation profile, and so the sign of the angular momentum deposited
changes at this depth.  At small depths (before the peak) the angular
momentum deposited has a sign such as to reduce $|\Delta\Omega|$.  The
angular momentum removed from this region gets deposited deeper in the
star over a distance which is of order a wave damping length (see Fig.
2).

Initially gravity waves force a shell of thickness equal to a
dissipation length of waves with $\omega \approx N_c$, lying just
below the convection zone, to corotate with the convection zone; this
occurs on a time scale of only $\sim 10^3$ years (fig. 3).  The
angular momentum removed from this region is deposited over the next
dissipation length. This process continues and the thickness of the
shell corotating with the convection zone grows with time.  Since the
dissipation length increases with depth, the angular momentum removed
from the corotating shell is redistributed over increasingly larger
distances.  These effects are all seen in Figure (2). On a timescale
$\approx 10^7$ years the size of the differentially rotating core of
the sun has shrunk to less than $\approx 0.5 R_\odot$.  Will this
process lead to the entire radiative interior rotating uniformly?  For
simplicity, we have assumed that waves of different $m$ are excited to
the same amplitude. This means that the net angular momentum
luminosity at the base of the convection zone is zero, and thus any
excess/deficit of angular momentum in the radiative interior (over
corotation with the base of the convection zone) gets concentrated in
a region near the center of the sun the size of which decreases with
time. Moreover, we have not considered the effect of waves which
return to the convection zone from the radiative interior; these waves
will be reabsorbed by the convection zone, thus removing the
excess/deficit of angular momentum from the core of the sun.  Thus, we
believe that allowing for a net exchange of angular momentum between
the convection zone and the radiative interior will lead to uniform
rotation of the entire radiative interior on the time scale of $\sim
10^7$ years estimated above.

\bigskip
\centerline{\bf 4. Discussion} 
\medskip

We have calculated the excitation of low frequency gravity waves by
turbulent convection in the sun and the angular momentum luminosity
they carry. The energy luminosity in dipole gravity waves with
frequencies of order $N_c \sim 0.15 \mu$Hz (the characteristic
convective frequency at the bottom of the convection zone) is
$\approx$ 10$^{29}$ erg s$^{-1}$ near the base of the convection
zone. Higher frequency dipole and quadrupole gravity waves 
are excited most efficiently by energy bearing turbulent eddies
deeper in the convection zone, and the resultant power spectrum is
$\omega^{-13/3}$. Waves of higher $\ell$ are more evanescent in the
convection zone and are thus less efficiently excited. Low frequency
gravity waves suffer strong radiative damping in the solar interior;
dipole waves with frequencies of $\approx$ 0.15$\mu$Hz travel only
$\sim 10$ wavelengths ($\approx 0.02 R_\odot$) into the radiative
interior, whereas $\ell=1$(10) waves of frequency 1.0$\mu$Hz have
dissipation lengths $\sim$ 1.0(0.1) $R_\odot$.
 
The angular momentum luminosity associated with a wave, i.e., the
amount of angular momentum crossing a sphere per unit time, is $m$
times the ratio of the energy luminosity and the wave frequency (where
$m$ is the azimuthal number of the wave).  The local damping rates for
$\pm m$ waves are in general different in a differentially rotating
medium, which results in a local deposition of angular momentum.  We
find that there is enough angular momentum in gravity waves generated
by convection that they can force much of the radiative interior of
the sun into corotation with the base of the convection zone in
$\approx$ 10$^7$ years (see Fig. 2 \& 3); Zahn et al. (1996) have 
independently arrived at a similar conclusion (see also, Schatzman 1996).

Our calculations were carried out under the assumption that waves of
different azimuthal number are excited to the same amplitude, as seen
in the local rest frame of the sun just below the convection
zone. Furthermore, we have neglected the effect of the coriolis force
on the wave function as well as the angular momentum removed from the
core of the sun by gravity waves which are reabsorbed by the
convection zone. As a consequence waves merely redistribute angular
momentum in the radiative interior of the sun in such a way that a
significant fraction of the radiative interior is forced into
corotation with the bottom of the convection zone, where the waves are
excited.  We feel that when these simplifying assumptions are dropped,
and a net exchange of angular momentum between the convection zone and
the radiative interior is allowed, the mechanism we have described
here will lead to uniform rotation of the entire radiative interior of
the sun.

\bigskip
\ni{\bf Acknowledgment:} While this work was in an advanced stage 
we found out that John-Paul Zahn was independently pursuing the same
idea of the angular momentum transport by gravity waves in stars.  PK
would like to thank him for many informative discussions, and for
pointing out the work of Schatzman. We would also like to thank Jeremy
Goodman and Ramesh Narayan for useful discussions. PK is grateful to
John Bahcall for encouraging him to pursue this work when he visited
IAS last year. The idea for this work was indirectly inspired by Peter
Goldreich's beautiful work on the synchronization of early type stars
in binary systems. This work was supported by NASA grant NAGW-3936.

\vfill\eject 

\centerline{\bf REFERENCES} 
\bigskip
\ni Balbus, S.A. and Hawley, J.F., 1996, to appear in MNRAS

\ni Christensen-Dalsgaard, J. et al. 1996, Science, 272, 1286

\ni Goldreich, P., Murray, N., and Kumar, P. 1994, ApJ 424, 466 

\ni Goldreich, P. \& Nicholson, P.D., 1989, ApJ 342, 1079

\ni Press, W.H. 1981, ApJ, 245, 286 

\ni Schatzman, E., 1996, J. Fluid Mech., 322, 355

\ni Thompson, M.J., et al. 1996, Science, 272, 1300

\ni Zahn, J-P, Talon, S. \& Matias, J., 1996, submitted to AA

\vfill\eject
\bigskip
\centerline{\bf Figure Captions}
\bigskip

\ni FIG. 1.--- (a)  The net energy luminosity in gravity waves 
($L_E$) as a function of radius in the sun. Wave generation in the convective 
overshoot layer is not included. (b) The frequency of
dipole waves, as a function of radius in the sun, for which the energy
luminosity in the wave has decreased by a factor of $e$ from its value
at the base of the convection zone (i.e., the wave's attenuation
depth, $\tau$, is $\approx 1$). These waves carry most of the energy
luminosity at the given radius.  We note that $\omega_{\tau = 1}
\propto [\ell(\ell+1)]^{3/8}$.  

\ni FIG. 2.--- (a) The time evolution of the differential rotation
profile of the sun, $\Delta \Omega(r)$, which is obtained by solving
eqs. (\ref3) \& (\ref2). The curve marked with $t=0$ is an arbitrarily
chosen initial rotation profile and is not intended to reflect the
sun's actual primordial rotation profile.  The energy luminosity in
gravity waves, needed for this calculation, is given in Figure (1).
Panel (b) shows the net angular momentum luminosity in gravity waves,
$L_a(r)$, at the three different epochs considered in panel (a).  

\ni FIG. 3.--- $t(r)$ is the characteristic time over which the solar 
rotation rate at radius $r$ changes due to the angular momentum
deposited by gravity waves. $t(r)$ is much less than the age of the
sun everywhere in the radiative interior and so the angular momentum
carried by gravity waves can strongly modify the solar rotation
profile.

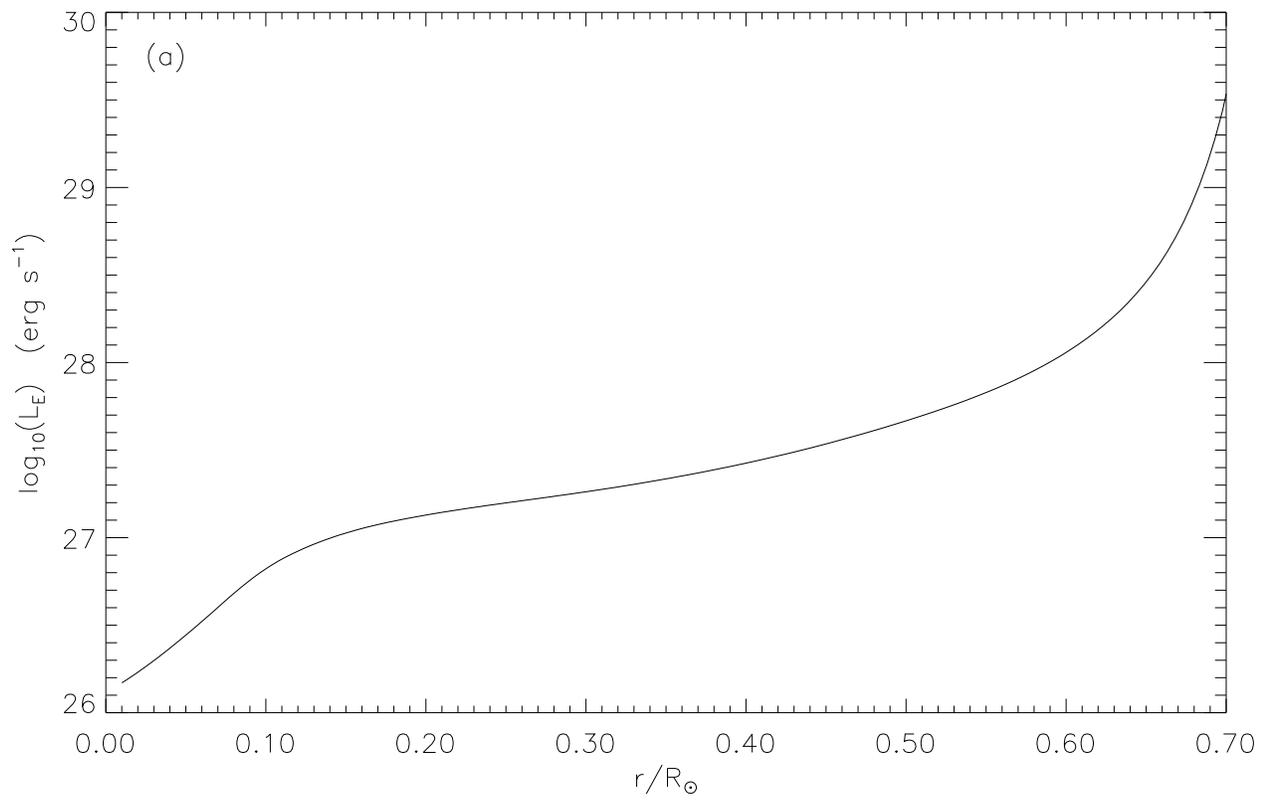
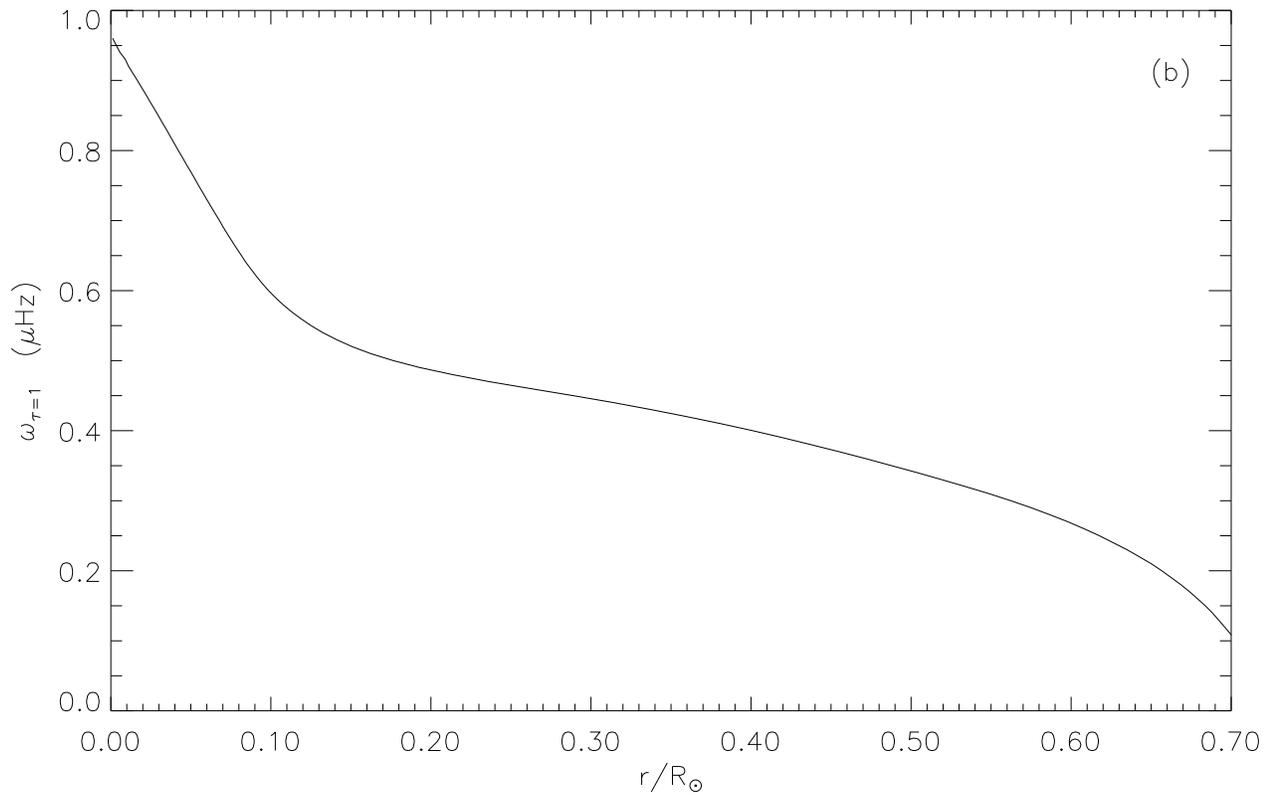
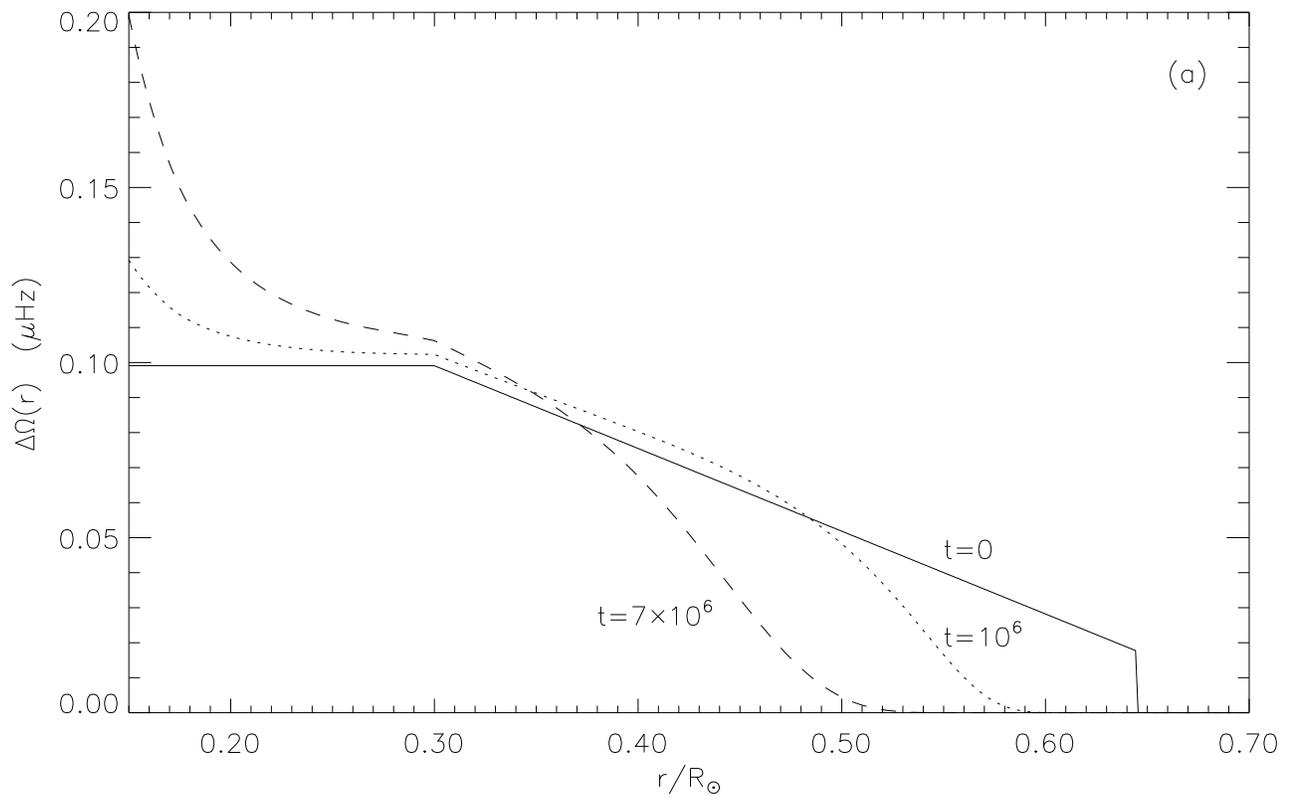
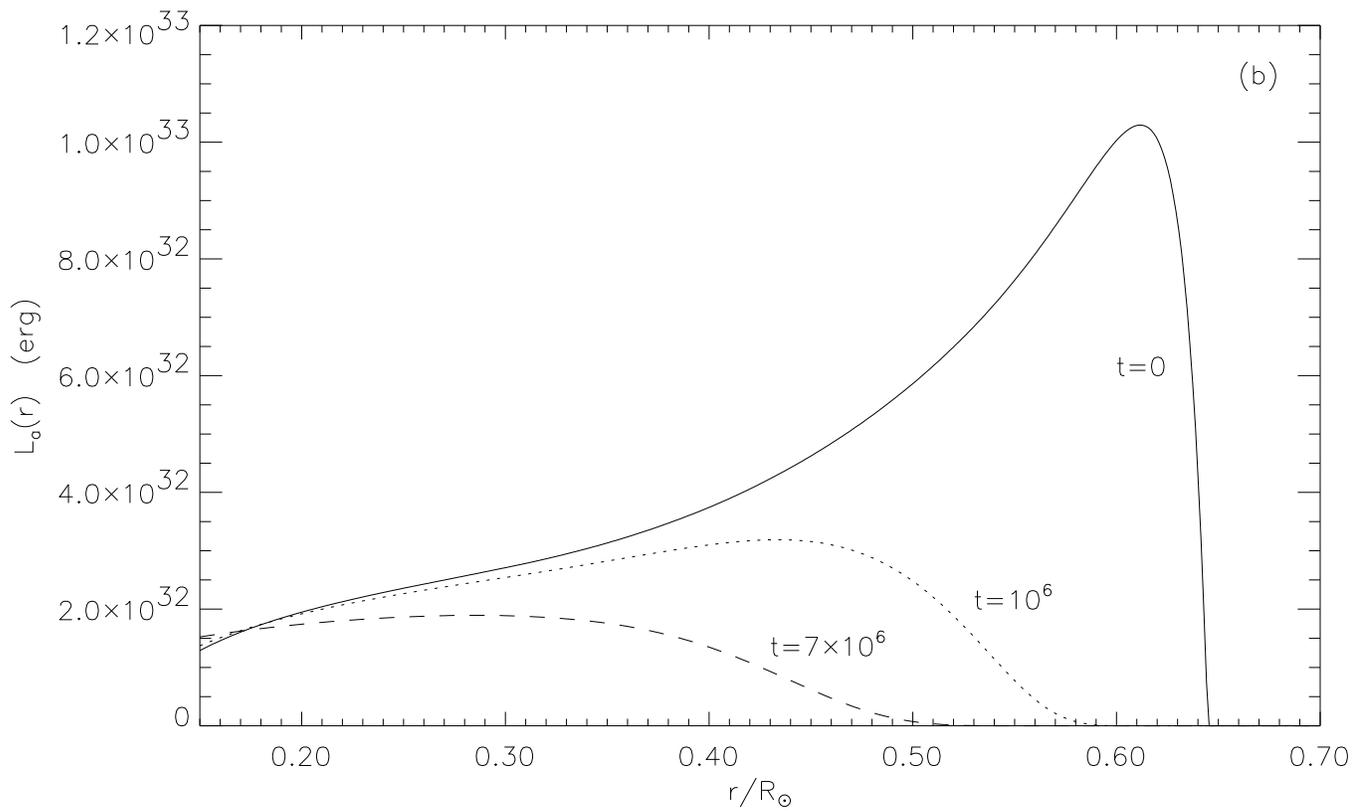
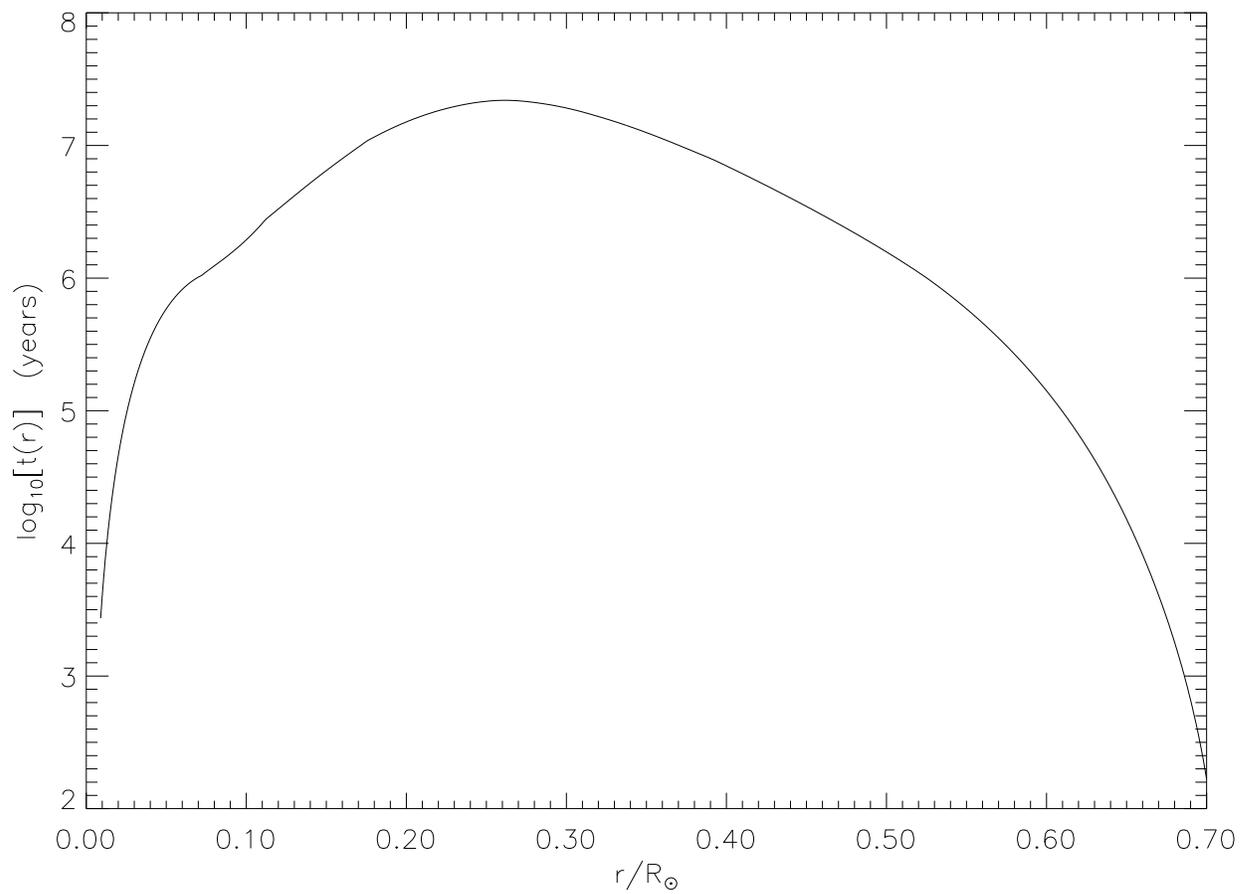
\bye